\documentclass[aps,prc,showpacs,nofootinbib]{revtex4}

\usepackage{epsfig}
\usepackage{graphicx}

\begin{document}

\title{On radiative corrections for unpolarized electron proton elastic scattering}

\author{Egle Tomasi-Gustafsson }

%\email{etomasi@cea.fr}
\affiliation{\it DAPNIA/SPhN, CEA/Saclay, 91191 Gif-sur-Yvette
Cedex, France }

\date{\today}
%\pacs{25.30.Bf, 13.40.-f, 13.40.Gp}

\begin{abstract}
A statistical analysis of the elastic unpolarized electron proton scattering data shows that, at large momentum transfer, the size and the $\epsilon$ dependence of the radiative corrections, as traditionally calculated and applied, may induce large correlations of the parameters of the Rosenbluth fit, which prevent a correct extraction of the electric proton form factor. Using the electron QED structure (radiation) function approach the cross section of elastic electron-proton scattering in leading and next-to leading approximations is calculated and expressed as a correction to the Born cross section, which is different for the electric and the magnetic contribution. When properly applied to the data, it may give the solution to the problem of the discrepancy of the polarized and unpolarized results on electron proton scattering.
\end{abstract}

\maketitle
%%%%%%%%%%%%%%%%%%%%%%%%%%%%%%%
\section{Introduction}
%%%%%%%%%%%%%%%%%%%%%%%%%%%%%%%

The experimental determination of the elastic proton electromagnetic form factors (FFs) at large momentum transfer is presently of large interest, due to the availability of electron beams in the GeV range with high intensity and high polarization, large acceptance spectrometers, hadron polarized targets, and hadron polarimeters. The possibility of extending the measurements of such fundamental quantities, which contain dynamical information on the nucleon structure, has inspired experimental programs at JLab, Frascati and at future machines, such as GSI, both in the space-like and in the time-like regions.

The traditional way to measure  proton electromagnetic FFs consists in the determination of the $\epsilon$ dependence of the reduced elastic differential cross section, which may be written, assuming that the interaction occurs through the exchange of one-photon, as \cite{Ro50}:
\begin{equation}
\sigma_{red}^{Born}(\theta,Q^2)=\epsilon(1+\tau)\left [1+2\displaystyle\frac{E}{m}\sin^2(\theta/2)\right ]\displaystyle\frac
{4 E^2\sin^4(\theta/2)}{\alpha^2\cos^2(\theta/2)}\displaystyle\frac{d\sigma}{d\Omega}=\tau G_M^2(Q^2)+\epsilon G_E^2(Q^2),
\label{eq:sigma}
\end{equation}
$$
\epsilon=[1+2(1+\tau)\tan^2(\theta/2)]^{-1},
$$
where $\alpha=1/137$, $\tau=Q^2/(4m^2)$, $Q^2$ is the momentum transfer squared, $m$ is the proton mass, $E$ and $\theta$ are the incident electron energy and the scattering angle of the outgoing electron, respectively, and $G_M(Q^2)$ and $G_E(Q^2)$ are the magnetic and the electric proton FFs and are functions of $Q^2$, only.  Measurements of the elastic differential cross section at different angles for a fixed value of $Q^2$ allow $G_E(Q^2)$ and $G_M(Q^2)$ to be determined as the slope and the intercept, respectively, from the linear $\epsilon$ dependence (\ref{eq:sigma}).

High precision data on the ratio of the electric to magnetic proton FFs at large $Q^2$ have been recently obtained  \cite{Jo00} through the polarization transfer method \cite{Re68}.  Such data revealed a surprising trend, which deviates from the expected scaling behavior previously obtained through the measurement of the elastic  $ep$ cross section according the Rosenbluth separation method  \cite{An94}. New precise measurements of the unpolarized elastic $ep$ cross section \cite{Ar04} and re-analysis of the old data \cite{Ch04,Ar04b} confirm that the behaviour of the measured ratio $R(Q^2)=\mu  G_E(Q^2)/G_M(Q^2)$ ($\mu=2.79$ is the magnetic moment of the proton) is different depending on the method used:
\begin{itemize}
\item  Scaling behavior for unpolarized cross section measurements: $R(Q^2)\simeq 1$; $G_M(Q^2)$ has been extracted up to $Q^2\simeq 31$ GeV$^2$ \cite{Ar75} and is often approximated, for practical purposes,  according to a dipole form: $G_D(Q^2)=(1+Q^2/0.71\mbox{~GeV}^2)^{-2}$;
\item a strong monotonical decrease from polarization transfer measurements.
\begin{equation}
R(Q^2)=1-(0.130\pm 0.005)\{Q^2~[\mbox{GeV}^2]-(0.04\pm 0.09)\}.
\label{eq:brash}
\end{equation}
\end{itemize}
The ratio deviates from unity as $Q^2$ increases, reaching a value of $\simeq$ 0.35 at $Q^2=$ 5 GeV$^2$ \cite{Jo00}.

This puzzle has given rise to many speculations and different interpretations \cite{Bl03,Gu03,Ch04b}, suggesting further experiments. In particular, it has been suggested that the presence of $2\gamma  $ exchange could solve this discrepancy through its interference with the main mechanism ($1\gamma$ exchange). In a previous paper \cite{ETG} it was shown that the present data do not give any evidence of the presence of the $2\gamma$ mechanism, in the limit of the experimental errors. The main reason is that, if one takes into account $C$-invariance and crossing symmetry, the $2\gamma$ mechanism introduces a very specific non linear $\epsilon$ dependence of the reduced cross section \cite{Re1,Re03t,Re04a}, whereas the data do not show any deviation from linearity.

Before analyzing the data in a different perspective, we stress the following points:
\begin{itemize}
\item No experimental bias has been found in both types of measurements, the experimental observables being the differential cross section on the one hand, and the polarization of the outgoing proton in the scattering plane (more precisely the ratio between the longitudinal and the transverse polarization), on the other hand.
\item The discrepancy is not at the level of these observables: it has been shown that constraining the ratio $R$ from polarization measurements and extracting $G_M(Q^2)$ from the measured cross section {\it "the magnetic FF is systematically 1.5-3\% larger than had been extracted in previous analysis" }, inside the error bars \cite{Br03}.

\item The inconsistency arises at the level of the slope of the $\epsilon$ dependence of the reduced cross section, which is  directly related to $G_E(Q^2)$, i.e. the derivative of the differential cross section, with respect to $\epsilon$. The difference of such slope, derived from the two methods above, appears particularly in the latest precise data \cite{Ar04}. One should note that the discrepancy appears in  the ratio $R$, whereas $G_M(Q^2)$  decreases more than one order of magnitude from $Q^2$=1 to 5 GeV$^2$.
\end{itemize}

Radiative corrections to the unpolarized cross section can reach 30-40\% at large $Q^2$. RC are calculated as a global factor which is applied to the number of detected elastic events, $\sigma^{meas}$. As a rule, they depend on the kinematical variables, as $\epsilon$ and $Q^2$. They are traditionally applied to the unpolarized differential cross section, following a prescription which includes only leading order contributions \cite{Mo69,Shwinger}:
\begin{equation}
\delta \simeq -\displaystyle\frac{2\alpha}{\pi}
\left \{ \ln \displaystyle\frac{\Delta E}{E}
\left [\ln \left (-\displaystyle\frac{q^2}{m^2}\right )-1\right ]+ 
\displaystyle\frac{3}{4}\ln \left (-\displaystyle\frac{q^2}{m^2}\right )+f(\theta)\right \}
\label{eq:eqts}
\end{equation}
where $f(\theta)$ is function only of the scattering angle $\theta$.

As noted in the original papers \cite{Mo69,Shwinger}, when $\Delta E\to 0$, 
$\sigma^{meas}$ becomes negatively infinite, whereas physical arguments require that it should vanish. The authors stated already that this problem would be overcome taking into account higher order radiative corrections. 

In recent experiments $E$ is large and the experimental resolution is very good  (allowing to reduce $\Delta E$).  Moreover, multiple photon emission from the initial electron, shifts the momentum transferred to the proton to lower values, increasing the cross section. Therefore $\delta$ becomes sizeable and one can not safely neglect higher order corrections. 

A complete calculation of radiative corrections should take into account consistently all different terms which contribute at all orders (including the two photon exchange contribution) and their interference.

However, several approximations are made, which may not be safely extrapolated to the conditions of the present esperiments. In particular in the calculation of Ref. \cite{Mo69}, the consideration of hard collinear photon emission (where the radiative photon is emitted along the direction of the incident or outgoing electron) is not complete. Moreover higher order RC, pair production as well as vacuum polarization are not included. 

We calculate here the cross section of elastic electron-proton scattering in leading and next-to leading approximation using the electron QED structure (radiation) function approach, which takes into account any number of real and virtual photons, emitted in collinear kinematics, at all orders in QED. This approach was previously applied to unpolarized $e^+e^-$ scattering \cite{Ku85}, to deep inelastic scattering \cite{Ku88} and, more recently, to polarization observables in $ep$ elastic scattering \cite{Af00,DKSV}. It was found that the correction is lower than 1\% to polarization observables, as expected. However, the effect on the polarized cross section was not investigated, and, in particular the effect on the slope of the reduced cross section as a function of $\epsilon$, which is the relevant quantity here.

The purpose of this paper is to re-analyze the unpolarized data, with particular attention to the applied RC. 

The paper is organized as follows. In Section II we show that a large correlation exists between the two parameters extracted from the Rosenbluth fit at large $Q^2$ and analyze the existing data in this respect. A probable source of these correlations being found in the standard procedure taken for RC, we calculate RC for the cross section of elastic electron-proton scattering in frame of the structure functions (SF) approach (Section III). In Section IV numerical results are presented, which show that the correction to the measured cross section is different for the electric and the magnetic contribution, and therefore affects the slope of the reduced cross section and, in particular, the extraction of $G_E(Q^2)$.

%%%%%%%%%%%%%%%%%%%%%%%%%%%%%%%%%%%%%%%%%%%%%%%%%%%%%%%%%
\section{Statistical analysis of the present data}
%%%%%%%%%%%%%%%%%%%%%%%%%%%%%%%%%%%%%%%%%%%%%%%%%%%%%%%%%%%%%%

\begin{figure}
\begin{center}
\includegraphics[width=17cm]{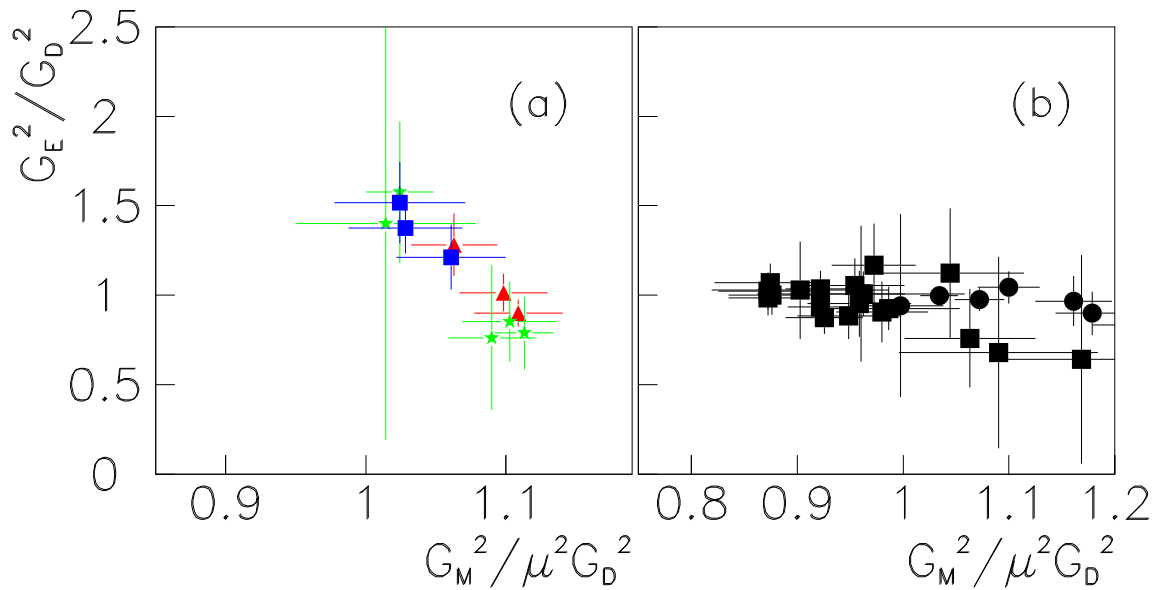}
\caption{\label{Fig:fig1} Dependence of  $G_E^2/G_D^2$ versus 
$G_M^2/\mu^2 G_D^2$:
(a) for $Q^2 \ge$ 2 GeV$^2$ from Refs. \protect\cite{Ar04} (triangles), 
\protect\cite{Ch04} (stars) and \protect\cite{Wa94} (squares); 
(b) for $Q^2 \le$ 2 GeV$^2$ from Refs. \protect\cite{Be71} (circles), 
and \protect\cite{Ja66} (squares). 
}
\end{center}
\end{figure}

The starting point of this work is the observation of a correlation, which appears in the published FFs data extracted with the Rosenbluth method: the larger is $G_E^2$, the smaller $G_M^2$. The dependence of $G_E^2/G_D^2$ versus $G_M^2/\mu^2 G_D^2$  is shown in Fig. \ref{Fig:fig1}a for three recent data sets, at $Q^2\ge$ 2 GeV$^2$\cite{Ar04,Ch04,Wa94}. In Fig. \ref{Fig:fig1}b two data sets at low $Q^2$ ($Q^2 \le$ 2 GeV$^2$) are shown \cite{Be71,Ja66}. Whereas at low $Q^2$, $G_E^2/G_D^2$ seems constant and quite independent from $G_M^2/\mu^2  G_D^2$, at large $Q^2$ an evident correlation appears. 
This is especially visible in the most recent and precise experiment, at large $Q^2$ \cite{Ar04}, where a linear fit of the ratio $R(Q^2)$ as a function of $Q^2$ gives:
\begin{equation}
R(Q^2)=(0.13\pm 0.11)\{Q^2~[\mbox{GeV}^2]+(0.57\pm 0.32)\}.
\label{eq:arri}
\end{equation}
where $Q^2$ is expressed in GeV$^2$.

Polarization data also showed a linearity of the ratio $R$, with the same slope (in absolute value), Eq. (\ref{eq:brash}), but with  opposite sign. In this case, the ratio is measured directly, whereas according to the Rosenbluth method one extracts two (independent) parameters from a linear fit. A correlation between the two parameters could be induced by the procedure itself or could be a physical effect and have a dynamical origin. In the latter case, it should not depend on the experiment. 

\begin{figure}
\begin{center}
\includegraphics[width=12cm]{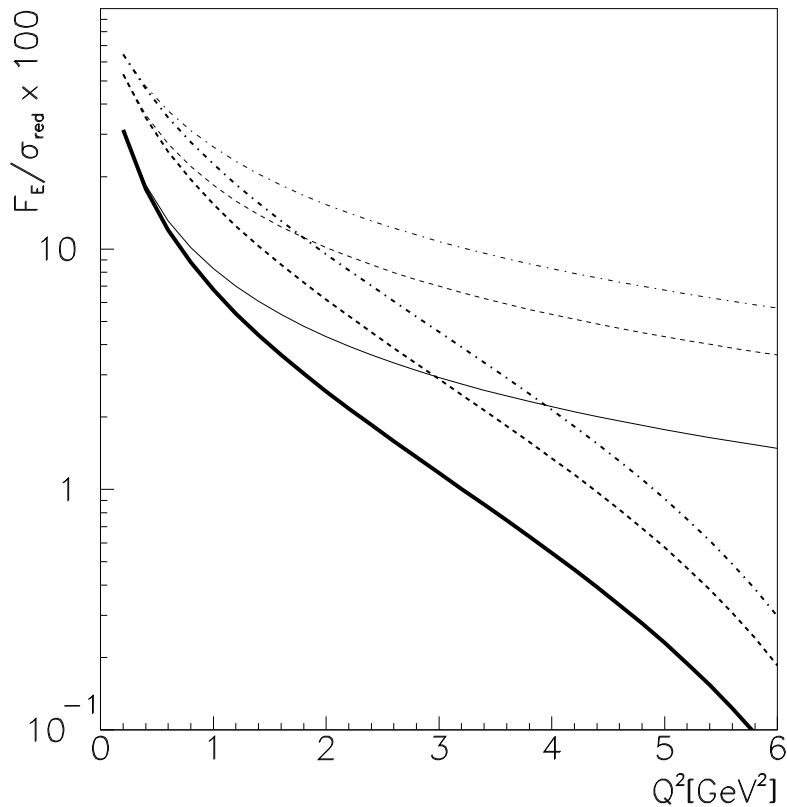}
\caption{\label{Fig:fig2} Contribution of the $G_E(Q^2)$ dependent term to the reduced cross section (in percent) for $\epsilon=0.2$ (solid line), $\epsilon=0.5$ (dashed line), $\epsilon=0.8$ (dash-dotted line), in the hypothesis of FF scaling (thin lines) or following Eq. (\protect\ref{eq:brash}) (thick lines).}
\end{center}
\end{figure}

It is known that at large $Q^2$ the contribution of the electric term to the cross section becomes very small, as the magnetic part is amplified by the kinematical factor $\tau $. This is illustrated in Fig. \ref{Fig:fig2}, where the ratio of the electric part, $F_E=\epsilon G_E^2(Q^2)$, to the reduced cross section is shown as a function of $Q^2$. The different curves correspond to different values of $\epsilon$, assuming FFs scaling (thin lines) or in the hypothesis of the linear dependence of Eq. (\ref{eq:brash}) (thick lines). In the second case, one can see that, for example, for $\epsilon=0.2$ the electric contribution becomes lower than $3$\% starting from 2 GeV$^2$. This number should be compared with the absolute uncertainty of the cross section measurement. When this contribution is larger or is of the same order, the sensitivity of the measurement to the electric term is lost and the extraction of $G_E(Q^2)$ becomes meaningless.
Since the first measurements \cite{Ho62}, electromagnetic probes have been traditionally preferred to hadronic beams, as the electromagnetic interaction is exactly calculable in QED, and one can safely extract the information from the hadronic vertex. However, one has to introduce the radiative corrections, which become very large as the momentum transfer squared, $Q^2$, increases. Radiative corrections were first calculated by Schwinger \cite{Shwinger} and are important for the discussion of the experimental determination of the differential cross section.

The measured elastic cross section is corrected by a global factor $C_R$, according to the prescription \cite{Mo69}:
\begin{equation}
\sigma_{red}^{Born}= C_R\sigma_{red}^{meas}.
\label{eq:sred}
\end{equation}
The factor $C_R$ {\it contains a large $\epsilon$ dependence} and a smooth $Q^2$ dependence,  and it is common to the electric and magnetic parts. At the largest $Q^2$ considered here this factor can reach 30-40\%, getting larger when the resolution is higher. If one made a linear approximation for the uncorrected data, one might even find a negative slope starting from $Q^2\ge 3$ GeV$^2$ \cite{ETG}.

\begin{figure}
\begin{center}
\includegraphics[width=15cm]{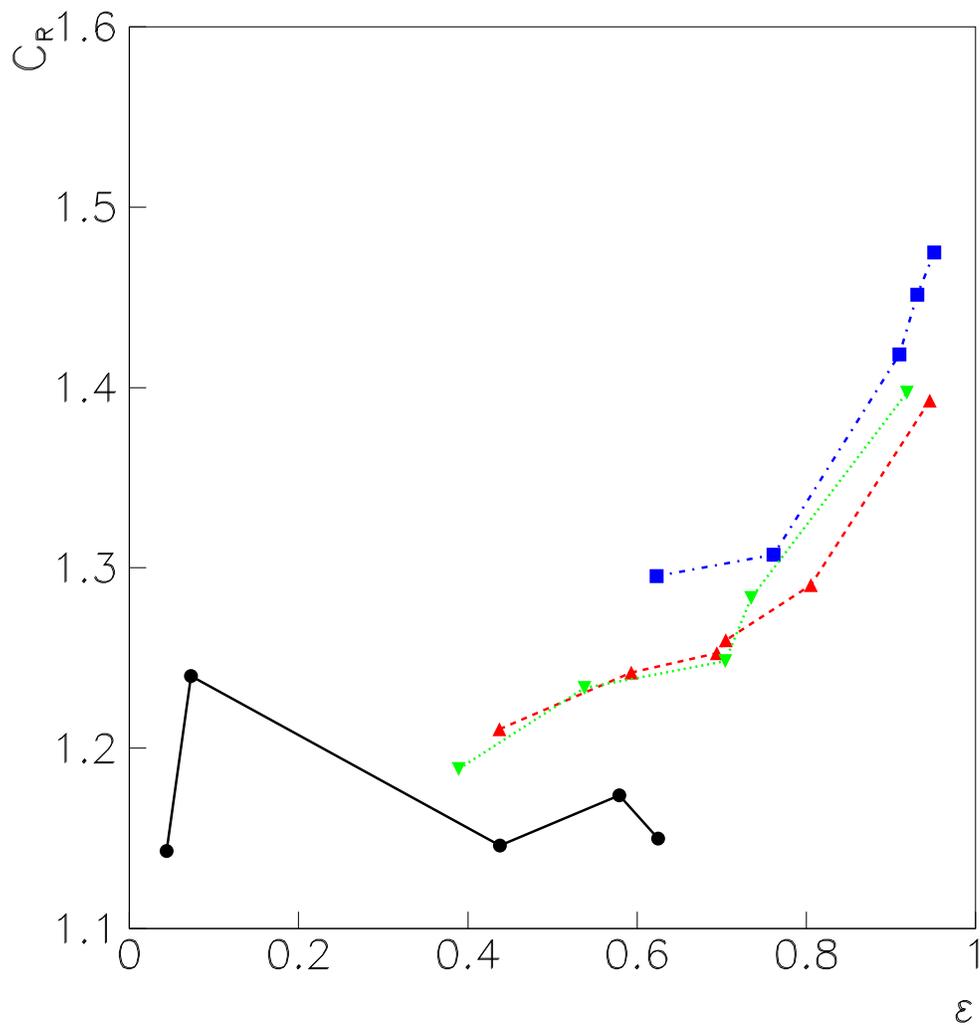}
\caption{\label{Fig:fig3} Radiative correction factor applied to the data at $Q^2$=3 GeV$^2$  (squares) from Ref. \protect\cite{Wa94}, at $Q^2$=4 GeV$^2$ (triangles) and 5 GeV$^2$ (reversed triangles) from Ref. \protect\cite{An94}, and at $Q^2$=0.32 GeV$^2$ from Ref. \protect\cite{Ja66} (circles). The lines are drawn to guide the eye. }
\end{center}
\end{figure}

In Fig. \ref{Fig:fig3} we show the $C_R$ dependence on $\epsilon$ for different $Q^2$ and from different sets of data. One can see that $C_R$ increases with $\epsilon$, rising very fast as $\epsilon\to 1$. It may be different in different experiments because its calculation requires an integration over the experimental acceptance.

The Rosenbluth separation consists of a linear fit to the reduced cross section at fixed $Q^2$, where the two parameters are $G_E^2$ and $ G_M^2$.
The multiplication by a factor which is common to the electric and magnetic terms, see Eqs. (\ref{eq:sigma},\ref{eq:sred}), and depends strongly on $\epsilon$, induces a correlation between these two parameters. In order to determine quantitatively how large this correlation is, we have built the error matrix for the Rosenbluth fits to the different sets of data available in the literature.

\begin{figure}
\begin{center}
\includegraphics[width=17cm]{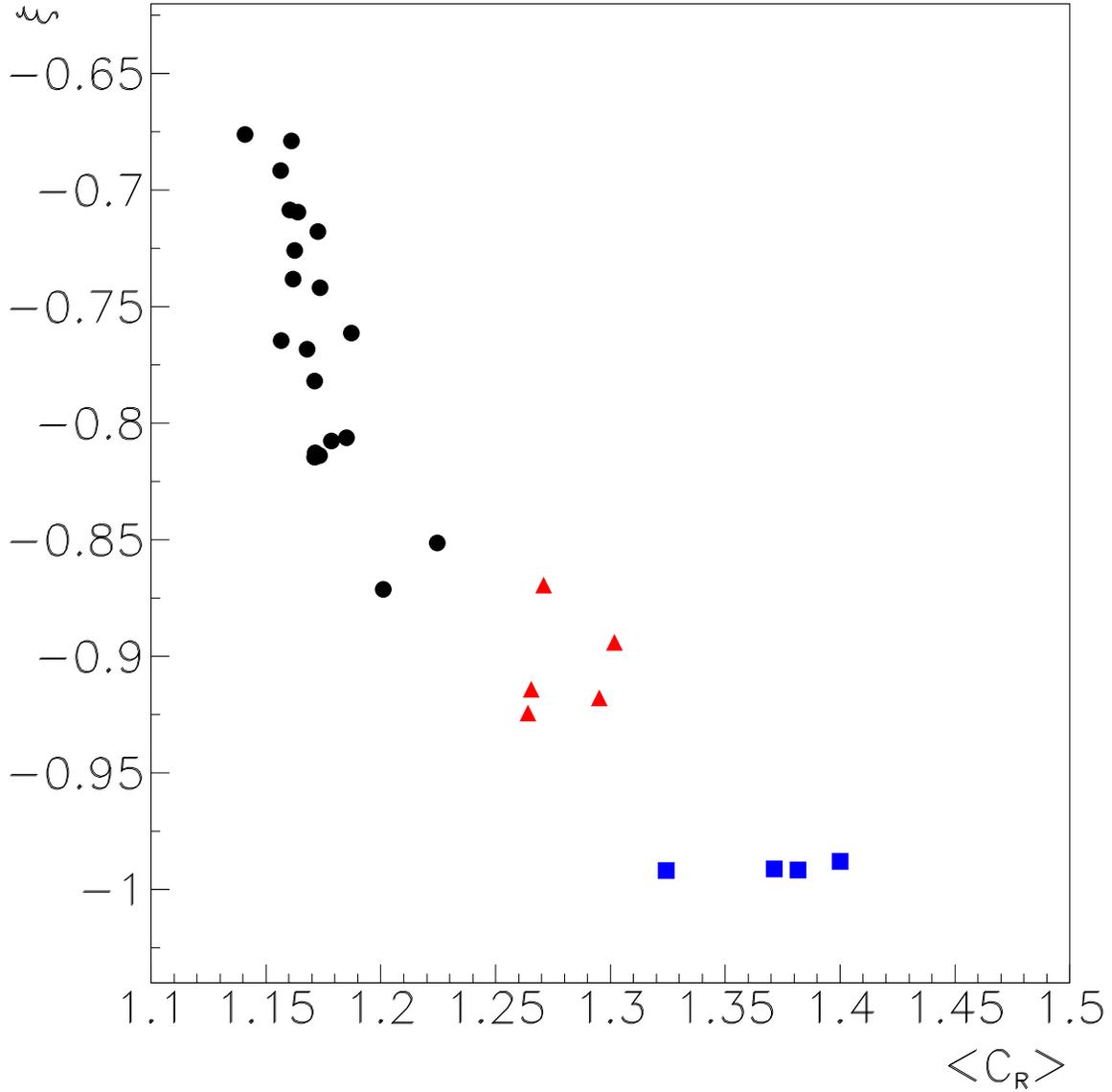}
\caption{\label{Fig:fig4} Correlation coefficient, $\xi$, as a function of the radiative correction factor $<C_R>$, averaged over $\epsilon$, for different sets of data: from Ref. \protect\cite{Ja66}  (circles), from Ref. \protect\cite{An94} (triangles) and from  Ref. \protect\cite{Wa94} (squares). }
\end{center}
\end{figure}

At fixed $Q^2$ the reduced cross section, normalized to $G_D^2$, has been parametrized by a linear $\epsilon$ dependence:
$\sigma_{red}^{Born}/G_D^2=a\epsilon +b$. The two parameters, $a$ and $b$, have been determined for each set of data as well as their errors $\sigma_a$, $\sigma_b$ and the covariance, $cov(a,b)$. The correlation coefficient $\xi$ is defined as
$\xi=cov(a,b)/\sigma_a\sigma_b$ and is shown in Fig. \ref{Fig:fig4} as a function of the average of the radiative correction factor $<C_R>$, weighted over $\epsilon$.

As the radiative corrections become larger, the correlation between the two parameters also becomes larger, reaching values near its maximum (in absolute value). Full correlation means that the two parameters are related through a constraint, i.e. it is possible to find a one-parameter description of the data. 
The data shown in Fig. \ref{Fig:fig4} correspond to those sets of experiments where the necessary information on the radiative corrections is available. The correlation coefficient itself can be calculated for a larger number of data and it is reported in Table \ref{table:table1}.
\newpage\clearpage
\begin{center}
\begin{table}[ht]
\begin{tabular}{|c|c|c||c|c|c|}
\hline
$Q^2$ [GeV] &$\xi$ & Ref. & $Q^2$ [GeV] &$\xi$ & Ref.\\
\hline
    2.6400 &	 -0.8823 & \protect\cite{Ar04} &      0.2717 &     -0.7258   & \protect\cite{Ja66}  \\ 
    3.2000 &	 -0.8973 &&  			       0.2911 &     -0.7818   & \\ 
    4.1000 &	 -0.9060 & &     		       0.3105 &     -0.7085   & \\ 
    1.7500 &	 -0.8693  &\protect\cite{An94} &      0.3493 &     -0.7683   & \\ 
    2.5000 &      -0.9141 &&   		       0.3881 &     -0.7417   & \\ 
    3.2500 &      -0.9242 && 			       0.4269 &     -0.7093   & \\ 
    4.0000 &	 -0.9178  && 			       0.4657 &     -0.7381   & \\ 
    5.0000 &	 -0.8940 & &       		       0.5045 &     -0.8126   & \\ 
    1.0000 &	 -0.9918 &\protect\cite{Wa94}  &    0.5433 &     -0.7646   & \\ 
    2.0030 &	 -0.9915  && 			       0.5821 &     -0.8076   & \\ 
    2.4970 &	 -0.9910  && 			       0.6209 &     -0.8061   & \\ 
    3.0070 &	 -0.9878 & &        		       0.6598 &     -0.8137   & \\ 
    0.1552 &	 -0.6761 &\protect\cite{Ja66}  &    0.6986 &     -0.8713   & \\ 
    0.1785 &	 -0.6788 & & 			       0.7374 &     -0.8145   & \\ 
    0.1940 &	 -0.6915 & & 			       0.7762 &     -0.8512   & \\ 
    0.2329 &	 -0.7177  & &  		       0.8538 &     -0.7612   & \\ 
\hline    
\hline
\end{tabular}
\caption[]{ Correlation coefficient $\xi=cov(a,b)/\sigma_a\sigma_b$ for different sets of data.} 
\label{table:table1}
\end{table}
\end{center}

At low $Q^2$ a correlation still exists, but it is smaller. For the data from Ref. \cite{Ja66} the radiative corrections are of the order of 15\%, seldom exceed 25\% and  correspond to  $\epsilon < 0.8$. This allows a safer extraction of the FFs.

Fig. \ref{Fig:fig4} shows that, for each $Q^2$, the extraction of FFs by a two parameter fit may be biased by the $\epsilon$ dependence induced by the radiative corrections. Whatever the precision of the individual measurements is, the slope of the reduced cross section is not sensitive to $G_E(Q^2)$ at large $Q^2$. The $Q^2$ dependence is therefore driven by $G_M(Q^2)$, which follows a dipole form. For each $Q^2$ a nonzero value of the ratio $R$  will lead to an apparent dipole dependence of $G_E(Q^2)$. Therefore experiments based on this method, will always give a $Q^2$ dependence of $G_E(Q^2)$ which is driven by  $G_M(Q^2)$, i.e. follow approximately a dipole behavior.

%%%%%%%%%%%%%%%%%%%%%%%%%%%%%%%%%%%%%%%%%%%%%%%%%%%%%%%%%%
\section{Calculation of radiative corrections}
%%%%%%%%%%%%%%%%%%%%%%%%%%%%%%%%%%%%%%%%%%%%%%%%%%%%%%%%%%%%

It is known \cite{Ku88} that the process of emission of hard photons by initial and scattered electrons plays a crucial role, which results in the presence of the radiative tail in the distribution on the scattered electron energy. We give here a simplified example of how a different calculation of radiative corrections can affect the electric and magnetic part of the unpolarized cross section, and change, in particular, its $\epsilon$ dependence. The main point of interest here is to show the very sharp dependence of the initial state emission on the inelastic tail of the scattered electron energy spectrum. A more extended version of this calculation and its application to polarization observables, including two photon exchange, is given elsewhere \cite{Ku06}. 

The aim of this paper is to drive the attention to the sensitivity of the electric FF to the procedure used for its extraction from the data, and to focus the attention on how radiative corrections are applied on the unpolarized cross section. 

The structure functions (SF) approach extends the traditional one \cite{Mo69}, taking precisely into account the contributions of higher orders of perturbation theory and the role of initial state photon emission. The cross section is expressed in terms of SF of the initial electron and of the fragmentation function of the scattered electron  energy fraction. Experimentally the detection of the scattered electron does not allow to separate the collinear photon emission. Therefore, one integrates in a range of the scattered electron energy. This is equivalent to set the fragmentation function to unity, due the well known properties of this formalism \cite{Ku88}.

It is known that initial state photon emission is more important than final state photon emission, due to the effect of decreasing $Q^2$. Proton emission (real and virtual) is essentially smaller than the electron ones, and can be included as a general normalization. Vacuum polarization, which has been often neglected in previous analysis, here is taken into account. 

The four momentum transfer squared can be written as: 
$$Q^2=\frac{2E^2(1-\cos\theta)}{\rho},$$
where $\rho$ is the recoil factor:
$$\rho=1+\frac{E}{M}(1-\cos\theta).$$
In an experiment, the selection of elastic scattering requires the integration of the events in the elastic peak, and the rejection of inelastic events. We parametrize the cut on the energy of the final electron $E'$, selecting events with $E'>c/\rho$, where $c$ is the  'inelasticity' cut, $ c<1$ (for the present numerical application we choose $c=0.97$).

Due to the properties of SF method, radiative corrections can be written in form of initial and final state emission, although gauge invariance is conserved. This form obeys the Lee-Nauenberg-Kinoshita theorem, about the cancellation of mass singularities, when integrating on the the final energy fraction. This results in omitting the final (fragmentation) SF, i.e., in replacing the structure function associated with the final electron emission by unity. 

Therefore, the differential cross section, calculated in frame of the SF method, $\frac{d\sigma^{SF}}{d\Omega}$, can be written as \cite{Ku05}:
\begin{equation} \frac{d\sigma^{SF}}{d\Omega}=\frac{\alpha^2\cos^2(\theta/2)}{4E^2\sin^4(\theta/2)}\int_{z_0}^1dzD(z)\frac{\phi(z)}{[1-\Pi(Q^2_z)]^2}\left (1+\displaystyle\frac{\alpha}{\pi}K\right ).
\label{eq:eqdz}
\end{equation} 
where $K$ is an $\epsilon$-independent quantity of the order of unity, which includes all the non-leading terms, as two photon exchange and soft photon emission. More precisely the interference between the two virtual photon exchange amplitude and the Born amplitude and the relevant part of the soft photon emission i.e., the interference between the electron and proton soft photon emission, are included in the term $K$. This effect is not enhanced by large logarithm (charcteristic of SF) and can be included in non-leading contributions. The factor $1+\frac{\alpha}{\pi }K$ can be considered as a general normalization. It is calculated in detail in Ref. \cite{Ku06}. 

Here we focuss on the $\epsilon$-dependence of the differential cross section. The SF calculation, Eq. (\ref{eq:sigma}), can be expressed in the form of a correction the Born reduced cross section (we omit RC of higher order):
\begin{equation} 
\sigma_{red}^{SF}= \sigma_{red}^{Born}(1+\Delta^{SF})
\label{eq:csa}
\end{equation} 
with 
$$\Delta^{SF}=
\displaystyle\frac{\alpha}{\pi}\left \{ \displaystyle\frac{2}{3}(L-\displaystyle\frac{5}{3})-
\displaystyle\frac{1}{2}(L-1)
\left [2\ln\left (\displaystyle\frac{1}{1-z_0}\right )-z_0
-\displaystyle\frac{z_0^2}{2}\right ]+\right .
$$
\begin{equation}
\left . \displaystyle\frac{1}{2}\rho(1+\tau)(L-1)\int_{z_0}^1
\displaystyle\frac{(1+z^2)dz}{1-z}
\left [\displaystyle\frac{\phi(z)}{[1-\Pi(z)]^2}-
\displaystyle\frac{\phi(1)}{[1-\Pi(1])^2}\right ] \right \},~L=\ln\frac{Q^2}{m_e^2},
\label{eq:cs}
\end{equation} 
$m_e$ is the electron mass. The structure (radiation) function $D(z)$ is
\begin{equation} D(z)=\frac{\beta}{2 }
\left[(1+\frac{3}{8}\beta)(1-z)^{\frac{\beta}{2}-1}-\frac{1}{2}(1+z)\right ]+O(\beta^2),~\beta=\frac{2\alpha}{\pi}\left [\ln\frac{Q^2}{m_e^2}-1\right ]. 
\label{eq:eq6}
\end{equation}
The lower limit of integration, $z_0$, is related to the 'inelasticity' cut, $c$, necessary to select the elastic data:
\begin{equation}
z_0=\frac{c}{\rho -c(\rho-1)},
\label{eq:eqz}
\end{equation}
The transfer momentum and recoil factor of the scattered electron after the  collinear photon emission are, respectively, $Q_z$ and $\rho_z$:   
\begin{equation}
Q_z^2=2E^2z^2(1-\cos\theta)/\rho_z;
~\rho_z=1+z\frac{E}{M}(1-\cos\theta).
\label{eq:eqqz}
\end{equation}
The kinematically corrected Born cross section for the scattered electron, $\phi(z)$, is:
\begin{equation}
\phi(z)=\frac{1}{\epsilon_z z^2\rho_z(1+\tau_z)}\sigma_{red}(z),
~\sigma_{red}(z)={\tau_z}G_M^2(Q_z^2)+\epsilon_z G_E^2(Q_z^2).
\label{eq:eqphiz}
\end{equation}
with 
\begin{equation}
\tau_z=\frac{Q^2_z}{4M^2},~\frac{1}{\epsilon_z}=1+2(1+\tau_z)\tan^2(\theta/2).
\label{eq:eqtauz}
\end{equation}
The vacuum  polarization for a  virtual photon with momentum $q$, $q^2=-Q^2<0$, is included as a factor $1/[1-\Pi(Q^2)]$. The main contribution to this term arises from the polarization of electron-positron vacuum:
\begin{equation}
\Pi(Q^2)=\frac{\alpha}{3\pi}\left [L-\frac{5}{3} \right ].
\label{eq:eqpi}
\end{equation}

The calculation requires a specific procedure for the integration of the SF  $D(z)$, which has a singularity at the upper limit of integration, Eq. (\ref{eq:eqdz}). 

\begin{figure}
\begin{center}
\includegraphics[width=14cm]{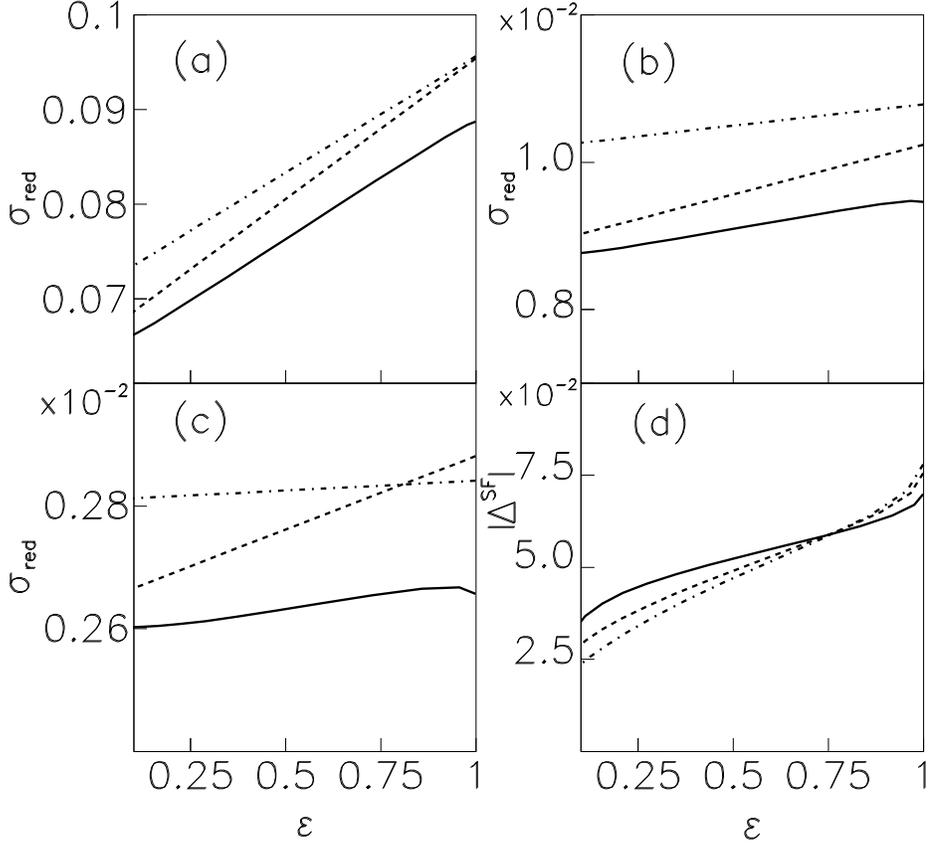}
\caption{\label{Fig:fig5}  Reduced cross section for $ep$ elastic scattering as a function of $\epsilon$, for $c=0.97$ at $Q^2$=1 GeV$^2$ (a), 3 GeV$^2$ (b), and  5 GeV$^2$ (b). The SF cross section, Eq. \protect\ref{eq:csa},  (solid line) and the Born cross section, Eq. \protect\ref{eq:sigma}, (dashed line) are shown for dipole parametrization of FFs. The absolute value ofthe correction, $|\Delta^{SF}|$, is shown in (d) for $Q^2$=1 GeV$^2$ (solid line), 3 GeV$^2$ (dotted line), and  5 GeV$^2$ (dash-dotted line). For comparison, the calculation of the Born cross section with FFs parametrized according to  \protect\cite{Br03} is shown as dash-dotted lines, in (a), (b) and (c). }
\end{center}
\end{figure}

The dependence of SF reduced cross section, Eqs. (\ref{eq:csa}-\ref{eq:eqpi}), on $\epsilon$ is shown in Figs. \ref{Fig:fig5}a,b,c, for different values of $ Q^2$=1, 3, 5 GeV$^2$, (solid lines). For comparison, the corresponding Born reduced cross section assuming also FFs parametrized in dipole form is shown as a dashed line, and the Born cross section, with FFs parametrized according to polarization measurements, (Eq. (\ref{eq:brash})  as a dash-dotted line.

One can see that $SF$ corrections affect the $\epsilon$ dependence of the cross section. Such effect is more important as $Q^2$ increases and for large  $\epsilon$ values. The relative difference of the SF reduced cross section with respect to the Born reduced cross section (both assuming dipole FFs), $|\Delta^{SF}|$, is shown in  Fig. \ref{Fig:fig5}d. For large values of $\epsilon$, the calculated reduced cross section can differ from the Born one by more than 7\%, for $c=0.97$. As both calculations assume dipole FFs, the source of the difference has to be attributed to how radiative corrections are calculated and applied.

Let us stress that the main effect of this correction is to modify and lower the slope of the reduced cross section. This effect brings into qualitative and quantitative agreement FFs data issued from polarized and unpolarized measurements, as one can see from the comparison of the solid and dash-dotted lines in Figs. \ref{Fig:fig5}a,b,c.

Of course, the concrete value of the slope depends on the inelasticity cut. Taking $0.95\le c\le 0.97$, the slope given by the SF calculation is in complete agreement with the slope suggested by the polarization measurements.

%%%%%%%%%%%%%%%%%%%%%%%%%%%%%%%%%
\section{Conclusions}
%%%%%%%%%%%%%%%%%%%%%%%%%%%%%%%%%%%
We reanalyzed the Rosenbluth data with particular attention to the radiative corrections applied to the measured cross section, and we showed from the (published) data themselves
that at large $Q^2$ statistical correlations between the parameters of the Rosenbluth plot become so large that $G_E(Q^2)$ can not be safely extracted. The method itself is biased at large momentum transfer because RC are applied as a global factor, which is the same for the electric and the magnetic contribution. Such factor contains a large $\epsilon$-dependence, which induces a strong correlation in the parameters of the linear $\epsilon$ fit. 

Calculations of RC in frame of the SF method, which takes into account higher order of perturbation theory, show that RC from collinear hard photon emission affect the elastic $ep$ cross section, in particular its $\epsilon$ dependence. Similarly to the standard RC, they depend on the electron scattering angle and on the kinematical selection for the elastic events. On the opposite, they act differently on the electric and magnetic term of the cross section, changing the slope of the reduced cross section which is related to the electric FF. When applied to the polarized cross section, it has been shown that their effect is small on the relevant observables \cite{Af00,DKSV}. Therefore it is suggested here that such corrections, when properly applied to the experimental data, can bring into agreement the results on the proton FFs issued from unpolarized and polarized measurements. Moreover these corrections affect very little the linearity of the Rosenbluth fit, contrary to what is expected from two photon exchange \cite{Re1}.

A complete calculation should take into account consistently all different terms which contribute at all orders (including the two photon exchange contribution) and their interference \cite{Ku06}. 

We confirm the conclusion of a previous paper \cite{Re68} which first suggested the polarization method for the determination of $G_E(Q^2)$, due to the increased sensitivity of the cross section to the magnetic term at large $Q^2$: '{\it Thus, there exist a number of polarization experiments which are more effective for determining the proton charge FF than is the measurement of the differential cross section for unpolarized particles}'.

\section{Acknowledgments}
This work was inspired by stimulating discussions with M. P. Rekalo. Thanks are due to J.L. Charvet, G.I. Gakh and  B. Tatischeff for useful suggestions and for a careful reading of the manuscript, to N. Keeley for improving the style of the manuscript. The structure function approach was proposed by E. A. Kuraev, who is acknowledged also for interesting lectures and discussions.

{}

\end{document}